\documentclass[12pt]{article}
\usepackage[utf8]{inputenc}
\usepackage[margin=0.8in]{geometry}
\usepackage{blindtext, xfrac}
\usepackage[english]{babel}
\usepackage[T1]{fontenc}
\usepackage{titling}
\usepackage{amsmath}
\usepackage{comment}
\usepackage{amssymb}
\usepackage{cite}
\usepackage{graphicx}
\usepackage{booktabs}
\usepackage{physics}
\usepackage{nameref}
\usepackage{ bbold }
\usepackage{textgreek}
\usepackage{hyperref}
\usepackage{gensymb}
\usepackage{float}
\usepackage[usenames,dvipsnames]{color}
\usepackage{svg}
\usepackage{placeins}
\usepackage{subcaption}
\usepackage{caption}
\usepackage{makecell}
\usepackage{authblk}

\newcommand{\Msun}{{\rm M}_\odot}

\title{High Mass Dark Matter Searches With the High Speed Flux From the Large Magellanic Cloud}

\author[1,2]{Nassim Bozorgnia}
\author[3,4,5]{Joseph Bramante}
\author[3,4]{Andrew Buchanan}

\affil[1]{Department of Physics, University of Alberta, Edmonton, Alberta, Canada}
\affil[2]{Theoretical Physics Institute, University of Alberta, Edmonton, Alberta, Canada}
\affil[3]{Arthur B. McDonald Canadian Astroparticle Physics Research Institute, 64 Bader Lane, Queen's University, Kingston, Ontario, Canada}
\affil[4]{Department of Physics, Engineering Physics, and Astronomy, Queen's University, Kingston, Ontario, Canada}
\affil[5]{Perimeter Institute for Theoretical Physics, Waterloo, Ontario, Canada}

\date{\today}

\begin{document}
\maketitle

\begin{abstract}
    As the hunt for dark matter progresses, recently there have been advances in the search for heavy dark matter with a mass well above a TeV. We show the importance of properly modeling the local dark matter velocity distribution, beyond the standard Maxwellian halo model, and in particular how the dynamics of the Large Magellanic Cloud and Milky Way may impact heavy dark matter searches. We introduce some new computational techniques for accurately computing the dark matter flux and the associated detector response. As a specific example, we examine the effect of the Large Magellanic Cloud on heavy dark matter bounds obtained from experiments searching for cosmic rays and magnetic monopoles using plastic etch detectors at the Ohya Mine and aboard the Skylab Space Station.
\end{abstract}

\section{Introduction}
For decades, astrophysicists have posited the existence of dark matter to explain several astronomical observations, including galactic rotation curves, the cosmic microwave background, gravitational lensing observations and galactic structure formation.
It has long been appreciated that the unit mass of dark matter in the galactic halo can be orders of magnitude larger than Standard Model particles \cite{Witten:1984rs,Farhi:1984qu,DeRujula:1984axn,Goodman:1984dc,Drukier:1986tm,Nussinov:1985xr,Bagnasco:1993st,Alves:2009nf,Kribs:2009fy}. Such massive states can be formed through dark sector nucleosynthesis \cite{Krnjaic:2014xza,Wise:2014jva,Gresham:2017cvl,Acevedo:2020avd}, dissipative processes in the early universe \cite{Buckley:2017ttd,Chang:2018bgx,Bramante:2023ddr,Bramante:2024pyc,Lu:2024xnb}, or a first order phase transition \cite{Witten:1984rs,Bai:2018dxf}. 
%Alternatively, the halo constituent need not be a fundamental particle and may be a large composite particle \cite{} or a gravitationally bound compact object. \cite{} %A bit more maybe 
Due to their low number density, such heavy halo constituents can have relatively high cross sections with Standard Model particles without conflicting with known bounds, and there are simple models that imply large cross sections for asymmetric composites via a weak coupling, for example a Higgs portal coupling to nucleons \cite{Acevedo:2021kly,Acevedo:2024lyr,Bleau:2025klr}. At large cross sections, dark matter can scatter multiple times with normal matter as it passes through the detector and overburden material. This means incoming dark matter may lose substantial energy to the overburden before reaching an underground detector, motivating searches nearer to the surface and even in outer space. Fortuitously, there are no known Standard Model background processes that leave a signature similar to that of such heavy dark matter, which deposits many nuclear recoils along a very straight trajectory. Because many solid-state multiscatter detectors require a minimum recoil energy per unit path length to register successive events, their sensitivity is especially dependent on the dark matter velocity distribution. In all prior work, multiscatter searches have assumed dark matter follows the Standard Halo Model (SHM).

The SHM~\cite{PhysRevD.33.3495} is the simplest and most commonly adopted model for the dark matter distribution in the Galaxy. In the SHM, the dark matter halo is assumed to be isothermal and spherical, with an isotropic Maxwell-Boltzmann velocity distribution which has a peak speed equal to the local circular speed. However, the actual dark matter distribution may have significant differences with respect to the SHM. In particular, high-resolution cosmological simulations that include both dark matter and baryons indicate that, while a Maxwellian distribution provides a reasonable fit to the local dark matter velocity distribution in simulated Milky Way (MW) analogues, the results carry large uncertainties~\cite{Bozorgnia:2016ogo, Kelso:2016qqj, Sloane:2016kyi,  Bozorgnia:2017brl,  Bozorgnia:2019mjk, Poole-McKenzie:2020dbo, Lawrence:2022niq, Kuhlen:2013tra, Lacroix:2020lhn, Santos-Santos:2023ubx, Herrera:2023fpq, Herrera:2021puj}.

One of the most important deviations from the SHM originates from the impact of the Large Magellanic Cloud (LMC) on the local dark matter distribution.
The LMC is the most massive satellite and the most recent merger of the MW. It has recently completed its first pericentric passage and is currently at a distance of $\sim 50$~kpc, moving at a speed of $321 \pm 24$~km/s  with respect to the MW~\cite{Besla:2007kf, Kallivayalil:2013xb,Patel:2017,Pietrzynski:2019}. Recent studies employing both cosmological~\cite{Smith-Orlik:2023kyl, Reynoso-Cordova:2024xqz} and idealized~\cite{Besla:2019xbx, Donaldson:2021byu} simulations demonstrate that the LMC significantly boosts the local dark matter velocity distribution to velocities above the Milky Way’s escape speed. This boost stems from the presence of high-speed dark matter particles originating from the LMC, as well as the gravitational influence of the LMC accelerating MW dark matter particles~\cite{Besla:2019xbx, Smith-Orlik:2023kyl}. As a result, the LMC can shift  direct detection exclusion limits by several orders of magnitude toward smaller dark matter-nucleon cross sections and lower dark matter masses, with the most pronounced impact for low-mass dark matter particles~\cite{Smith-Orlik:2023kyl, Reynoso-Cordova:2024xqz}.

Previous work \cite{Jacobs:2014yca,Bhoonah2021} computed exclusion limits for heavy dark matter, where \cite{Bhoonah2021} particularly used the SHM in its analysis and identified two existing searches for massive particles, namely the Skylab cosmic ray search and a search for monopoles at the Ohya quarry. The remainder of this paper is structured as follows. Section~\ref{sec:sims} describes the simulated MW-LMC analogue used in this work and our procedure for selecting a Solar neighborhood consistent with the observed LMC-Sun geometry. %, and the resulting local dark matter phase space distribution. 
In Section~\ref{sec:models} we lay out the multiscatter detector response formalism, including interaction models (spin-independent nucleon and contact cross sections), energy loss in overburden, directional efficiencies, and in Section \ref{sec:flux} we use discrete simulation samples to compute fluxes and energy thresholds for Ohya and Skylab. We then present the resultant limits in Section~\ref{sec:bounds}, comparing MW-LMC predictions to the SHM, and compare these to other searches. We conclude by summarizing the impact of the LMC-induced high-speed velocity tail on heavy-dark matter searches, discussing latitude-dependent flux effects, and outlining applications to future ground and orbital multiscatter detectors.

%
%{\bf things to add: 1) put in the params of the SHM. 2
%) put in a figure with the SHM model plotted alongside the binned LMC + MW model. 3) binned N$>v_0$ plot using real data as well + zoom in 4) for existing figure, label each sub-plot individually. 5) try to write a brief background for heavy dm and the point of this paper in the introduction. 6) say explicitly that the sums in eqs 22 and 23 are done by taking each velocity vector and integrating over the average of its passage. keep eq 22 and eq 23, but between the two write some explanatory text, that the first number is the total flux, while the second number is the flux above the velocity required to trigger the detector7) citation for galactic to equatorial coordinates8) Fix interpolation and bounds plots}

\section{Simulations}
\label{sec:sims}

%We now turn to our modeling of the local dark matter phase space distribution using an Auriga simulation MW–LMC analogue.

In this section we discuss the simulated MW–LMC analogue that we use to extract the  local dark matter phase space distribution and the precedure to specify the Solar region.

\subsection{MW-LMC analogue}

We use a MW-LMC analogue from the Auriga magneto-hydrodynamical simulations~\cite{Grand:2016mgo, Grand:2024}. The Auriga project~\cite{Grand:2016mgo} contains a set of cosmological zoom-in simulations of isolated MW-mass halos, selected from a (100~Mpc)$^3$ periodic cube (L100N1504) from the EAGLE project~\cite{Schaye:2014tpa, Crain:2015poa}. The simulations were carried out using the moving-mesh code Arepo~\cite{Springel:2009aa} and implement a galaxy formation subgrid model which includes star formation, supernova feedback, active galactic nuclei, black hole formation, metal cooling,  and background UV/X-ray photoionisation radiation~\cite{Grand:2016mgo}. The simulations adopt the Planck-2015~\cite{Planck:2015fie} cosmological parameters: $\Omega_{m}=0.307$, $\Omega_{\rm bar}=0.048$, $H_0=67.77~{\rm km~s^{-1}~Mpc^{-1}}$. We use the standard resolution level (Level 4) of the simulations, in which the dark matter particle mass is $m_{\rm DM} \sim 3\times 10^5~\Msun$, the baryonic particle mass is $m_b=5\times10^4~\Msun$, and the Plummer equivalent gravitational softening length is $\epsilon=370$~pc~\cite{Power:2002sw,Jenkins2013}.  The Auriga simulations reproduce the observed stellar masses, sizes, rotation curves, star formation rates and metallicities of present day MW-mass galaxies.

 The LMC's first pericenter approach occurred $\sim$ 50 Myr ago \cite{Besla:2007kf} at a distance of $\sim 48$~kpc~\cite{Besla:2007kf}, and the present-day stellar mass of the LMC from observations is $\sim 2.7 \times 10^9$~M$_\odot$~\cite{vanderMarel:2002kq}. We use a simulated MW-LMC analogue whose properties match those of the observed system. The analogue is the re-simulated halo 13 in ref.~\cite{Smith-Orlik:2023kyl}, corresponding to the Auriga 25 MW halo and its LMC analogue. This system was selected from a sample of 15 MW-LMC analogues in ref.~\cite{Smith-Orlik:2023kyl}, which were identified based on the present-day stellar mass of the LMC and its distance from host at first pericenter. Due to the large average time ($\sim 150$~Myr) between simulation snapshots in the 15 MW-LMC analogues, it is difficult to accurately identify their present-day snapshot. Therefore, one system (halo 13) was chosen and re-simulated in ref.~\cite{Smith-Orlik:2023kyl}, such that the average time between snapshots close to the LMC's pericenter approach is $\sim 10$~Myr, and the present-day snapshot closely matches observations. For this system, the virial mass of the MW host halo is $1.2 \times 10^{12}~\Msun$ and the halo mass of the LMC analogue at infall is $3.2 \times 10^{11}~\Msun$. More details on this system are provided in ref.~\cite{Smith-Orlik:2023kyl}. Finally, we note that we use galactocentric coordinates defined such that the DM wind will come from +90 degrees in longitude \cite{Bozorgnia:2016ogo,Smith-Orlik:2023kyl}. We find our directional component agrees with prior work, which sometimes uses the opposite convention for the galactocentric longitudinal coordinates \cite{Donaldson:2021byu}.

 %Two snapshots are considered for this system: isolated MW (\emph{Iso.}) and the present day MW-LMC (\emph{Pres.}). Iso.~occurs at the LMC analogue's first apocenter before infall when the MW and LMC analogues have the largest distance. Pres.~is the snapshot closest to the present-day separation of the observed MW and LMC. At the Pres.~snapshot, the distance of the LMC analogue from the host MW analogue is $\sim 50$~kpc, while its speed is 317~km/s. These closely match  their corresponding  values from observations. 

In this work, we utilize the present day MW-LMC snapshot for this system. This is the snapshot closest to the present-day separation of the observed MW and LMC, where the distance of the LMC analogue from the host MW analogue is $\sim 50$~kpc and its speed is 317~km/s. These closely match  their corresponding  values from observations. 

\subsection{Solar region}

There is no Solar position in the simulations a priori, and this needs to be specified. As discussed in  ref.~\cite{Smith-Orlik:2023kyl}, we choose the Sun's position and velocity in the simulated halo such that it matches the observed Sun-LMC geometry. This matching is performed in three steps. In the first step, we determine the orientations of the stellar disk in the MW analogue that reproduce the observed angle with the orbital plane of the LMC analogue. In the second step, for each allowed disk orientation we determine the Sun's position by requiring that the  angles between the LMC's orbital angular momentum and the Sun's position and velocity vectors in the simulations match their observed values. In the third step, we determine the best fit Sun's position as the one that yields the closest match of the angles between the Sun's velocity vector and the LMC's position and velocities with their observed values.

The \emph{Solar region} is defined as the overlap of two regions: a spherical shell with radii between 6 to 10 kpc from the MW analogue's center (with the Sun at a galactocentric distance of $\sim 8$~kpc), and a cone with its axis aligned with the best fit Sun's position, vertex at the galactic center, and an opening angle of $\pi/4$ radians. As detailed in ref.~\cite{Smith-Orlik:2023kyl}, the size of the Solar region is chosen such that it would be sensitive to the best fit Sun's position and also include several thousand dark matter particles. %Note that the Iso.~snapshot does not include an LMC analogue, and thus we cannot define the best fit Sun's position for this snapshot. Hence, for Iso. the Solar region is defined as the spherical shell with radii between 6 to 10~kpc from the center of the MW analogue.

\section{Heavy Dark Matter and Plastic Etch Detectors}
\label{sec:models}
We now lay out our treatment of heavy dark matter scattering in solid state detectors, and how this treatment is adapted to account for the high velocity component of dark matter associated with the LMC. Following \cite{Bhoonah2021}, we consider two different models of dark matter interactions with nuclei. The first is a spin-independent nuclear scattering cross section. In this case, omitting nuclear form factors, the dark matter-nuclear cross section is defined as
\begin{equation}
    \sigma_{\chi A} = A^{2} \frac{\mu_{\chi A}^{2}}{\mu_{\chi n}^{2}} \sigma_{\chi n},
\end{equation}
where $A$ is the number of nucleons in a nucleus, $\mu_{\chi A}$ is the dark matter-nucleus reduced mass, $\mu_{\chi n}$ is the dark matter-nucleon reduced mass and $\sigma_{\chi n}$ is the dark matter-nucleon cross section.
This interaction is appropriate if the dark matter is a loosely-bound composite, so that each dark matter constituent interacts coherently with all nucleons in a nucleus.
 The second model is a contact interaction, where the cross section is independent of the nucleus,
 \begin{equation}
    \sigma_{\chi A} = \sigma_{C}.
\end{equation}
This reproduces the classical result for a dark matter particle much larger than a nuclei and is appropriate for dark matter which is a single particle or a tightly-bound composite.

Next, we model the number of particles passing through the detector.
At any point in time, we model a plastic etch detector as a planar detector with a normal vector $\hat{n}(t)$ pointing out of the detector and perpendicular to the plane of the detector and a directional efficiency $ \epsilon(\hat{v},\hat{n}(t))$,
\begin{equation} \label{eq:EffFac}
   \epsilon(\hat{v},\hat{n}(t)) = \begin{cases}
1 & \text{if } \hat{n}(t) \cdot \hat{v}\leq-\cos(\theta_{\mathrm{tol}}), \\
0 & \text{if } \hat{n}(t) \cdot \hat{v} >-\cos(\theta_{\mathrm{tol}}),
\end{cases}
\end{equation}
where we note that since these are expressed in terms of a dot product between the detector normal vector $\hat{n}(t)$ and the dark matter velocity $\hat{v}$, this quantity is Euclidean frame-independent. Here, $\theta_{\mathrm{tol}}$ is the tolerance angle of the detector. 
The number of particles above a given speed, $v_0$, which pass through the detector is,
\begin{equation}\label{eq:speedFlux}
    N_{\geq v_0}(m_{\chi})= \frac{A_dT\rho_{\chi}}{m_{\chi}} \int_{v>v_0} d^3\mathbf{v} \  f(\mathbf{v}) I(\mathbf{v}),
\end{equation}
where $\rho_{\chi}=0.3\ \mathrm{ GeV}/\mathrm{cm}^3$ is the local dark matter density \cite{Eilers_2019dmspeed}, $m_{\chi}$ is the dark matter mass, $A_d$ is the area of the detector, and $T$ is the observation period for the detector. $f(\mathbf{v})$ is the local dark matter velocity distribution in the reference frame of the earth and $I(\mathbf{v})$ is a time average over detector quantities,
\begin{equation}
    I(\mathbf{v})=-\frac{1}{T}\int_0^{T} dt \ \ \hat{n}(t) \cdot \mathbf{v}\ \epsilon( \hat{v},\hat{n}(t) ) .
\end{equation}
This is a convenient formalism because it separates the dark matter and detector information, as $I(\mathbf{v})$ depends entirely on the properties of the detector and its orientation over time. 
In particular, it will be different for the Skylab and Ohya detectors. Setting $v_0=0$ gives the total number of particles passing through the detector,
\begin{equation}\label{eq:totalFlux}
    N (m_{\chi})= \frac{AT\rho_{\chi}}{m_{\chi}} \int d^3\mathbf{v} \  f(\mathbf{v}) I(\mathbf{v}),
\end{equation}
where we integrate over the entire velocity distribution.

We now describe the energetics of the system. As a dark matter particle crosses through a medium $M$, it will lose energy to background nuclei at a rate of,
\begin{equation}
    \frac{dE}{dx} = -\frac{2E}{m_\chi} \sum_{A\in M} \frac{\mu_{\chi A}^2}{m_A} n_A \sigma_{\chi A},
\end{equation}
where one sums over all nuclei in the material, labeled with mass number $A$ and with number density $n_A$.
It is helpful to rewrite the energy deposition rate in terms of an effective number density $\tilde{n}$, which depends on the medium and the dark matter mass,
    \begin{equation} \label{eq:energyLoss}
    \frac{dE}{dx} = -\tilde{n}_M\sigma E,
\end{equation}
where $\sigma$ is the cross section to be constrained. For a per-nucleon interaction, $\sigma = \sigma_{\chi n}$ and,
\begin{equation} \label{eq:densPN}
    \tilde{n}_{M}  = 2\sum_{A\in M} \frac{A^2\mu_{\chi A}^4}{m_\chi m_A\mu_{\chi n}^2} n_A.
\end{equation}
For a contact interaction, $\sigma = \sigma_{C}$ and,
\begin{equation}\label{eq:densCON}
    \tilde{n}_{M}  = 2\sum_{A\in M} \frac{\mu_{\chi A}^2}{m_\chi m_A} n_A.
\end{equation}

We now describe the bounds, which are computed using the same method as \cite{Bhoonah2021}. The experiments detected no straight line tracks through the detectors consistent with dark matter. To be conservative, we assume that the Ohya and Skylab experiments were unlucky and that tracks due to dark matter would have been observed 90\% of the time. This sets the average number of observed dark matter particles to be $N_c = -\ln(0.1)=2.3$, assuming Poisson statistics. The maximum dark matter mass an experiment excludes can be found by matching the number of observed dark matter particles, $N_c$, with the total expected number passing through the detector, $N(m_\chi)$. Next, to compute upper and lower bounds on the cross section, we determine the dark matter cutoff speed $v_0$ above which one sees $N_c$ particles pass through the detector, though not necessarily triggering it. We do so by solving,
\begin{equation}
    N_c = N_{\geq v_0}(m_{\chi}).
\end{equation}
Then, we take the initial dark matter energy to conservatively be $E_0=\frac{1}{2}mv_0^2$. The incoming dark matter must have enough energy to puncture the detector entirely. In this case, the energy deposition rate after the particle has passed through the detector must exceed the detector's energy threshold. Using Equation \ref{eq:energyLoss} and setting $E'_{\mathrm{th}}$ to be the energy threshold for the plastic track detector (measured in units of energy per unit length), one gets,
\begin{equation}\label{eq:enegyexponential}
    E'_{\mathrm{th}} \leq  \tilde{n}_DE_0\sigma \exp\left[-\sigma\int_{O+D} \tilde{n}(x)  dx\right],
\end{equation}
where we integrate over the path of the dark matter as it passes though the overburden and the entirety of the detector. We assume the detector has a constant density and constant isotope ratio through the particle's path. We also assume the overburden has a solid component with constant density and isotopic ratio as well as a potential atmospheric component. This gives,
\begin{equation}\label{eq:exponentialSup}
    E'_{\mathrm{th}} \leq  \tilde{n}_DE_0\sigma \exp\left[-(x_D\tilde{n}_D+x_S\tilde{n}_S+\tilde{\mathcal{N}}_A)\sigma\right].
\end{equation}
Here, $\tilde{n}_D$ is the effective number density of the detector, given in Equation \ref{eq:densPN} or \ref{eq:densCON}, and $x_D$ is length of the dark matter particle's path through the detector. $\tilde{n}_S$ and $x_S$ are defined similarly for the solid overburden. $\tilde{\mathcal{N}}_A=\int\tilde{n}_Adx$ is the effective column density of the atmosphere along the particles path. To be conservative, we assume the dark matter will traverse the overburden and hit the detector at the largest possible angle: the cutoff angle $\theta_{\mathrm{tol}}$. In Equation \ref{eq:exponentialSup}, $\sigma$ denotes either the per-nucleon cross section $\sigma_{\chi n}$ or the contact cross section $\sigma_{C}$.  

Equality in  Equation \ref{eq:exponentialSup} is satisfied by two values of the cross section $\sigma$. These solutions give the upper and lower bounds of the excluded region. They can be found numerically or written analytically with the transcendental Lambert-$W$ function:
\begin{align}\label{eq:bounds}
\begin{split}   
    \sigma_{\mathrm{upper}}&=-\frac{1}{x_D\tilde{n}_D+x_S\tilde{n}_S+\tilde{\mathcal{N}}_A}W_{-1}\left(-\frac{2(x_D\tilde{n}_D+x_S\tilde{n}_S + \tilde{\mathcal{N}}_A)E'_{\mathrm{th}} }{\tilde{n}_Dm_\chi v_0^2} \right),\\
        \sigma_{\mathrm{lower}}&=-\frac{1}{x_D\tilde{n}_D+x_S\tilde{n}_S+\tilde{\mathcal{N}}_A}W_{0}\left(-\frac{2(x_D\tilde{n}_D+x_S\tilde{n}_S+\tilde{\mathcal{N}}_A )E'_{\mathrm{th}} }{\tilde{n}_Dm_\chi v_0^2} \right),
\end{split}
\end{align}
where the $W_k$ are branches of the Lambert-$W$ function. 

For comparison, we will compute the bounds using both the SHM and the local dark matter velocity distribution extracted from the simulated MW-LMC analogue. In galactic coordinates\footnote{Here referring to the orientation of the axes, while the origin is placed on the Earth. The x-axis points toward the centre of the galaxy and the y-axis aligns with the direction of the Sun's orbital motion.}, we take the local velocity distribution in the Earth frame  under the SHM to be,

\begin{equation}\label{eq:shm}
    f(\mathbf{v}) \propto \exp\left(-\frac{\|\mathbf{v}-v_s\hat{y}\|^2}{v_\chi^2}\right)\Theta(v_{\rm esc}-\|\mathbf{v}-v_s\hat{y}\|),
\end{equation}
where $\Theta$ is the Heaviside step function. We take the escape speed of the MW to be $v_{\rm esc} = 503\: \text{km/s}$\cite{Deason2019}, the circular velocity at the Sun's position to be $v_\chi=222 \: \text{km/s}$\cite{Eilers_2019dmspeed} and the sun's speed relative to the galaxy to be $v_s=232 \:\text{km/s}$\cite{Springel:2009aa}. For both the SHM and the LMC data, we neglect the Earth's motion relative to the sun when computing bounds. The dark matter velocities from the simulations have been scaled by (220 km/s)$/v_{\chi}^{\rm sim}$, where $v_{\chi}^{\rm sim}$ is the local circular speed of the simulated Milky Way analogue. This allows us to compare the results from the simulated MW halo with the SHM, assuming a similar local circular speed.

\begin{figure}
    \centering
    \includegraphics[width=\linewidth]{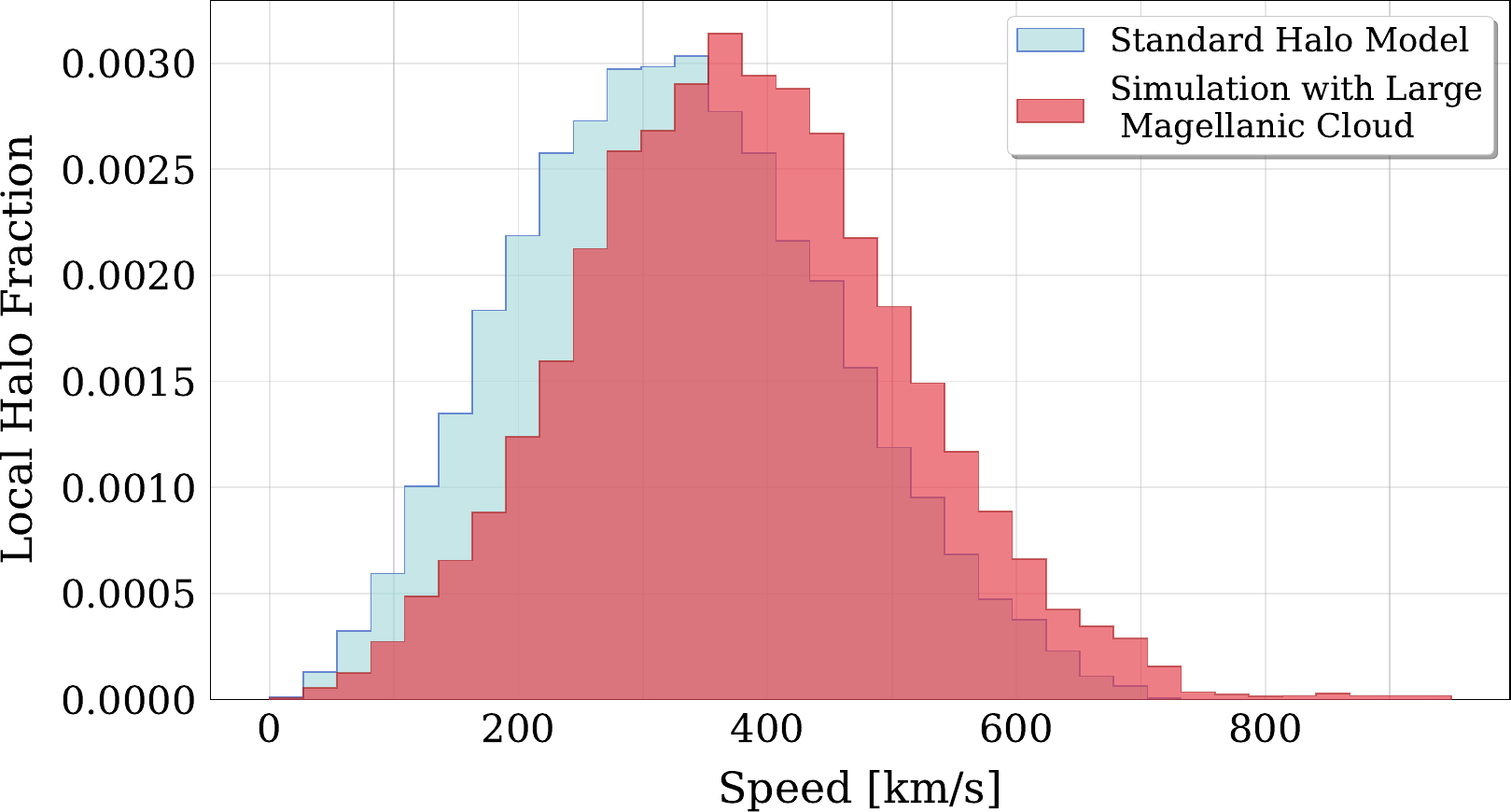}
    \caption{A comparison of dark matter's local speed distribution in the Earth reference frame for the SHM (blue) and the simulated MW analogue including the effect of the LMC (red).}
    \label{fig:histogram}
\end{figure}

\begin{figure}[ht]
    \centering
    \begin{subfigure}{0.85\textwidth}
        \centering
        \includegraphics[width=\textwidth]{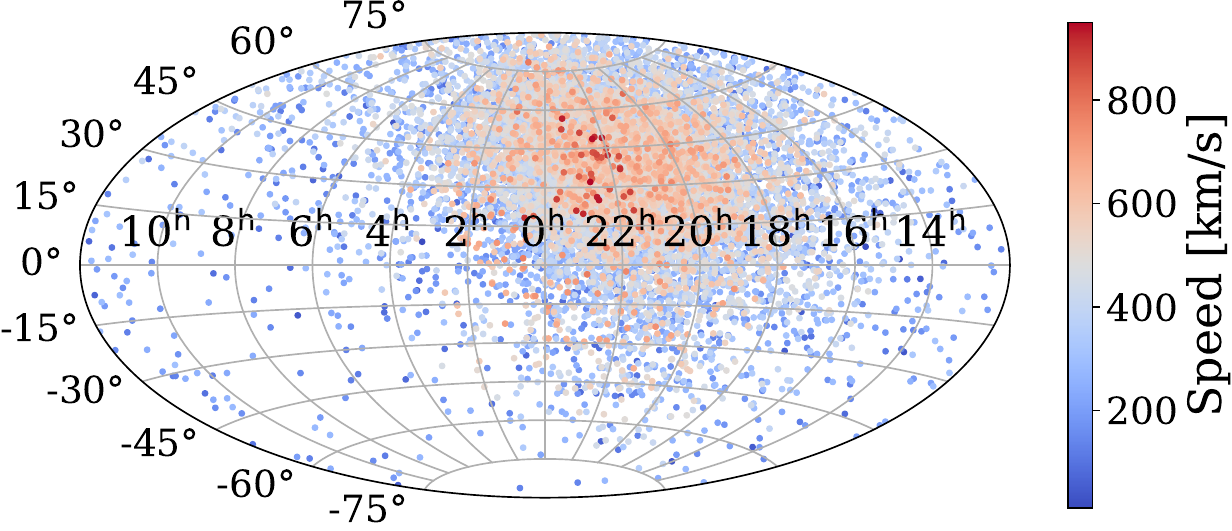}
    \end{subfigure}
    \hfill
    \begin{subfigure}{0.85\textwidth}
        \centering
        \includegraphics[width=\textwidth]{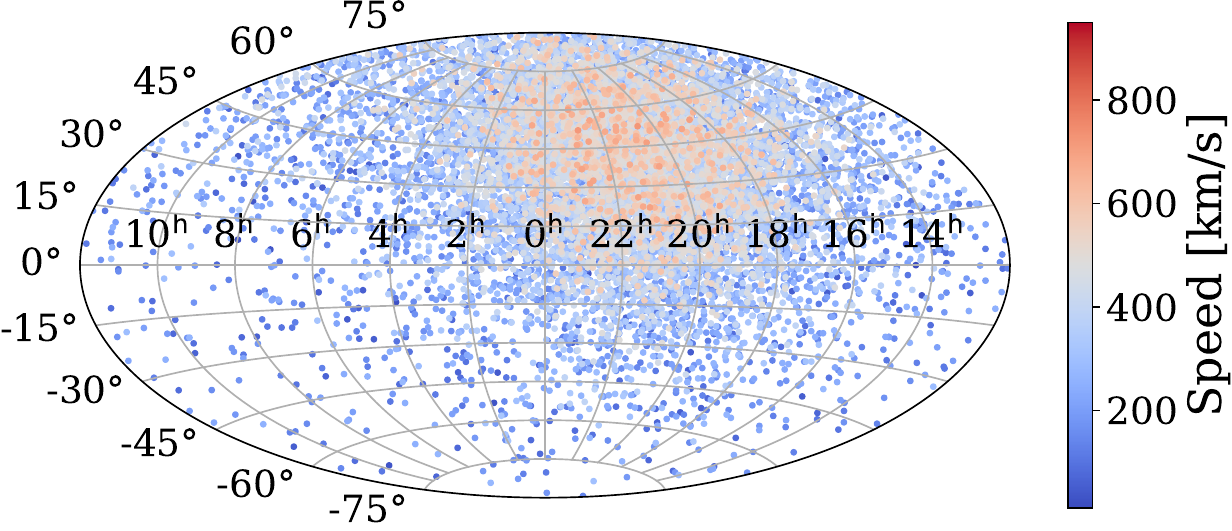}
    \end{subfigure}
    \centering
    \caption{Comparison of the speed and direction of origin for incoming dark matter particles assuming the results of the MW-LMC simulation (top) and the SHM (bottom). The samples are normalized to have approximately the same number of particles. The axes are the declination and right ascension of the incoming particle. The color bars specify the speed of the incoming dark matter particles in the Earth reference frame.}\label{fig:Direction Scatter}
\end{figure}

In Figures \ref{fig:histogram} and \ref{fig:Direction Scatter}, we compare the local dark matter distribution predicted by the SHM to that extracted from the simulations. Figure \ref{fig:histogram} compares their Earth frame speed distributions and Figure \ref{fig:Direction Scatter} shows the scatter plots of the incoming dark matter particles by direction for the two cases. Notice that, regardless of the model, in Figure \ref{fig:Direction Scatter} most of the fast incoming particles have declinations between $15\degree$ to $45\degree$. This is due to the Sun's orbital motion through the galaxy as it moves toward a point in the sky with $\approx 30 \degree$ declination. Figures \ref{fig:histogram} and \ref{fig:Direction Scatter} show that, in the Earth frame, the effect of the LMC is both to increase the speed of incoming dark matter and to concentrate the incoming dark matter in a smaller region of the sky. The latter effect, in particular, motivates a careful three dimensional treatment of the incoming dark matter flux.


\section{Modelling the Flux}\label{sec:flux}
We proceed to computing the time average of the directional efficiency, given in Equation \ref{eq:EffFac}, for the Ohya and Skylab detectors. In particular, we discuss how we treat the results from the MW-LMC simulated halo, where the local dark matter velocity distribution is extracted from a discrete sample of velocities. %result of a simulation, which estimates the local velocity distribution with a discrete sample of velocities.

The Ohya detector is flush with the Earth's surface, meaning its orientation oscillates daily in the galactic reference frame. Since the observation takes place over many days ($T=2.1\ \mathrm{ yrs}$), we can approximate $I(\mathbf{v})$ as the average over a sidereal day,
\begin{equation}\label{eq:sideint}
    I(\mathbf{v})\approx -\frac{1}{T_{\rm side}}\int_0^{T_{\rm side}} dt \ \ \hat{n}(t) \cdot \mathbf{v}\ \epsilon(\hat{v},t ) .
\end{equation}

It is easiest to evaluate this integral in equatorial coordinates, which are galactic coordinates rotated so that the $x-y$ plane aligns with the equatorial plane. In these coordinates, the normal vector pointing out of the Ohya detector can be parametrized as, 
\begin{equation}
    \hat{n}(t)=\begin{bmatrix}
        \cos(\phi_O)\cos(\frac{2\pi t}{T_{\rm side}})\\
        \cos(\phi_O)\sin(\frac{2\pi t}{T_{\rm side}})\\
        \sin(\phi_O)
    \end{bmatrix},
\end{equation}
where $\phi_O=36.6 \degree$ is the latitude of the Ohya quarry.
We also write the incoming velocity vector in spherical coordinates as, 
\begin{equation}
    \hat{v}=-\begin{bmatrix}
        v\cos(\delta_\chi)\cos(\alpha_\chi)\\
        v\cos(\delta_\chi)\sin(\alpha_\chi)\\
        v\sin(\delta_\chi)
    \end{bmatrix},
\end{equation}
where $\delta_\chi$ and $\alpha_\chi$ are the declination and right ascension of the incoming dark matter in the sky.
In this case, the integral in Equation \ref{eq:sideint}  can be calculated analytically, 
\begin{equation}
    I(\mathbf{v})=\frac{v}{\pi}(\sin(\delta_\chi)\sin(\phi_O)\theta_{\mathrm{tol}}+\cos(\delta_\chi)\cos(\phi_O)\sin(\theta_{\mathrm{tol}})),
\end{equation}
where we define the angle,
\begin{equation}\label{eq:groundFlux}
    \theta_{\mathrm{max}} = \begin{cases}
\pi & \text{if } \pi - |\phi_O+\delta_\chi|<\theta_{\mathrm{tol}}, \\
\arccos\left(\frac{\cos(\theta_{\mathrm{tol}})-\sin(\delta_\chi)\sin(\phi_O)}{\cos(\delta_\chi)\cos(\phi_O)}\right) & \text{if } |\phi_O-\delta_\chi|\leq \theta_{\mathrm{tol}} \leq\pi - |\phi_O+\delta_\chi|, \\
0 & \text{if } \theta_{\mathrm{tol}}<|\phi_O-\delta_\chi|.
\end{cases}
\end{equation}

We note an interesting consequence of this formula. By placing a detector at a latitude matching the declinations of the fastest incoming particles, one can maximize the flux through the detector. In Figure \ref{fig:Direction Scatter}, one can see that most of the fast incoming particles have declinations of $15\degree$ to $45\degree$. Coincidentally, this aligns well with Ohya's latitude of $36.6 \degree$, meaning the Ohya detector was well positioned to maximize the dark matter flux despite being intended as a magnetic monopole search.

To compute $I(\mathbf{v})$ for the plastic etch detector on the Skylab satellite, we would need detailed information about the orientation of the satellite over time and the location of the plastic etch detectors in Skylab. Unfortunately, the Sklyab experiment was ran in the 1970s and the cassette tapes recording the orientation of the satellite are not accessible.\footnote{We thank the NASA Space Science Data Coordinated Archive (NSSDCA) for their assistance.} So, we will assume that the Skylab detector was oriented randomly over all directions. Since the tolerance angle of Skylab was large ($60\degree$), this is not an unreasonable assumption. So, the time average $I(\mathbf{v})$ is simply a uniform average over all directions $\hat{n}$,
\begin{equation}
    I(\mathbf{v})=\frac{1}{4\pi}\int_{|\hat{n}|=1} d\Omega \ \ \hat{n} \cdot \mathbf{v} \ \epsilon( \hat{v}, \hat{n} ) .
\end{equation}
This integral can also be computed analytically and the resulting efficiency factor is
\begin{equation}
I(\mathbf{v})=\frac{v\sin^2(\theta_{\mathrm{tol}})}{4} .
\end{equation}

Using these formulae, the integrals in Equations \ref{eq:speedFlux} and \ref{eq:totalFlux} can be directly computed when assuming an analytic velocity distribution like the SHM in Equation \ref{eq:shm}. However, from the simulations, we obtain 
a discrete set of velocities for the local dark matter particles. To perform the integrals in this case, we will treat the simulated data as if it were a random sample pulled from the true velocity distribution $f(\mathbf{v})$, meaning one can approximate $f(\mathbf{v})\approx \frac{1}{M}\sum_{i=1}^M\delta^3(\mathbf{v}-\mathbf{v}_i)$, where $M$ is the total number of simulated data points. So, the integral in Equations \ref{eq:speedFlux} and \ref{eq:totalFlux} become a sum over the $M$ discrete data points. We compute the total flux through the detector as
\begin{equation}\label{eq:fluxDiscrete}
N(m_{\chi})= \frac{AT\rho_{\chi}}{Mm_{\chi}} \sum_{i=1}^M I(\mathbf{v}_i).
\end{equation}
We can use this value of $N(m_\chi)$ to compute the high mass boundary of the bound.
Similarly, we compute the flux through the detector above a given astronomical cutoff speed $v_0$,
\begin{equation}\label{eq:fluxAboveSpeed}
    N_{\geq v_0}(m_{\chi})= \frac{AT\rho_{\chi}}{Mm_{\chi}} \sum_{i=1}^M \Theta(v_i-v_0)  I(\mathbf{v}_i),
\end{equation}
where $\Theta$ is the Heaviside step function. 

This allows us to compute the flux using the data from the simulated MW-LMC system, and it also provides a straightforward way to compute the flux for the SHM. Rather than directly integrating in Equations \ref{eq:speedFlux} and \ref{eq:totalFlux}, one can apply Equations \ref{eq:fluxDiscrete} and \ref{eq:fluxAboveSpeed} to an arbitrarily large sample from the velocity distribution.

\begin{figure}[t]
    \centering
    \begin{subfigure}{0.48\textwidth}
        \centering
        \includegraphics[width=\textwidth]{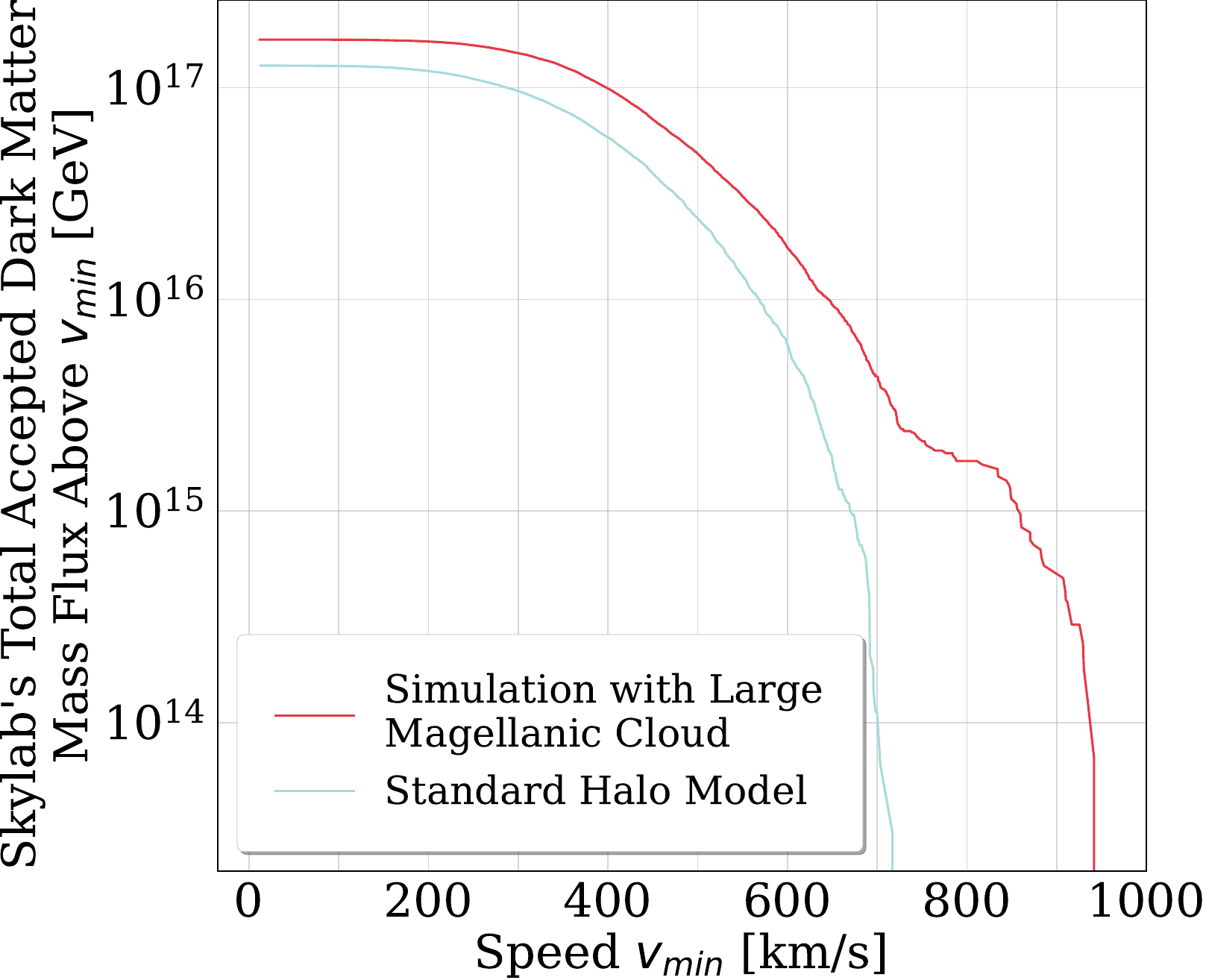}
    \end{subfigure}
    \hfill
    \begin{subfigure}{0.48\textwidth}
        \centering
        \includegraphics[width=\textwidth]{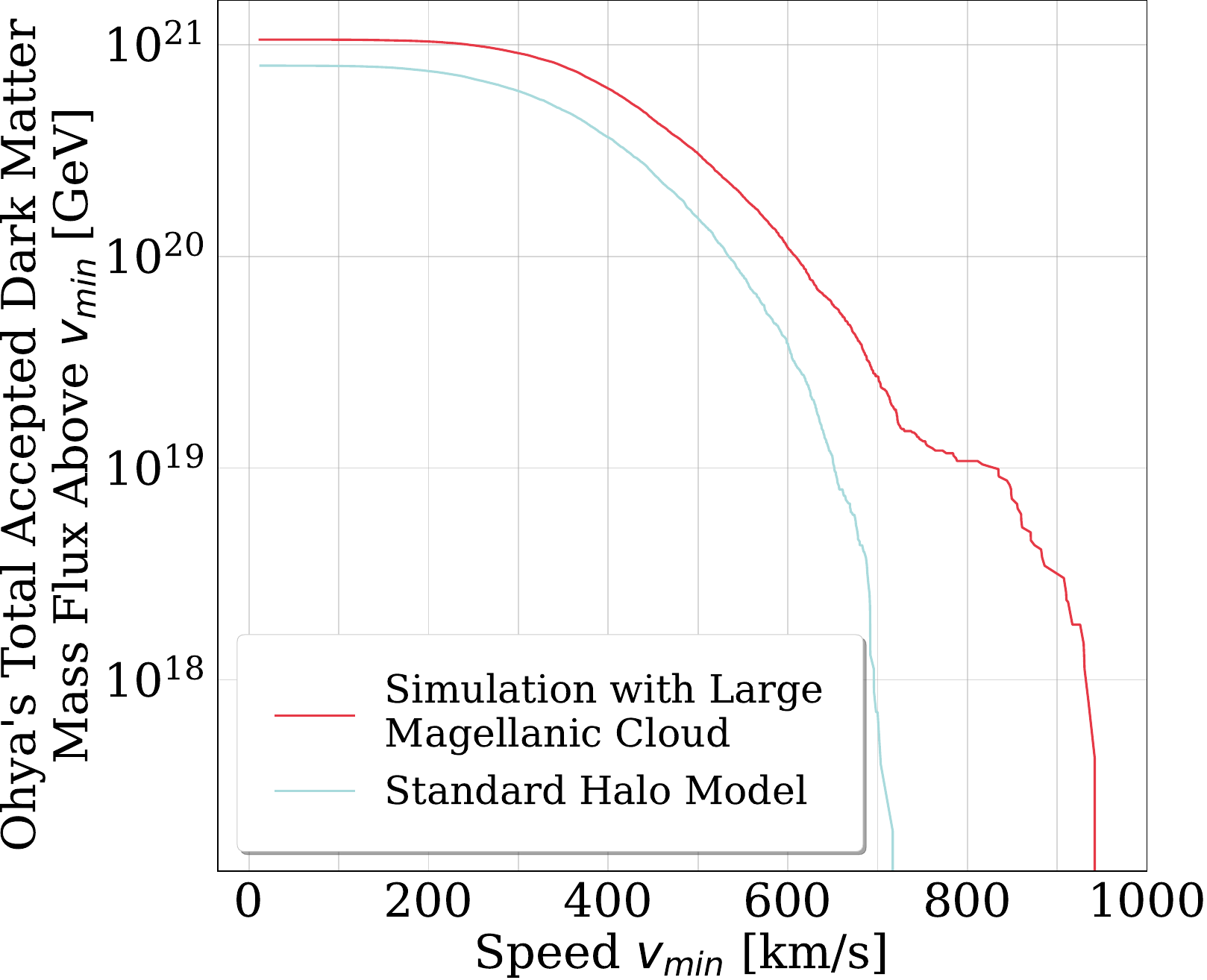}
    \end{subfigure}
    \centering
    \caption{Comparisons of the integrated dark matter mass flux intercepting the Skylab (left) and Ohya (right) detectors above a cutoff speed $v_{\mathrm{min}}$, between the SHM and the simulated MW analogue including the LMC. Note the significantly increased higher flux at high speeds for the simulated system compared to the SHM.}\label{fig:massFlux}
\end{figure}

Figure \ref{fig:massFlux} compares the integrated mass flux hitting the Ohya and Skylab detectors above a chosen speed, given by Equation \ref{eq:fluxAboveSpeed}, for both the MW-LMC simulation and the SHM. As expected, the presence of the LMC significantly increases the dark matter flux at high speeds.

\begin{figure}
    \centering
    \includegraphics[width=0.85\linewidth]{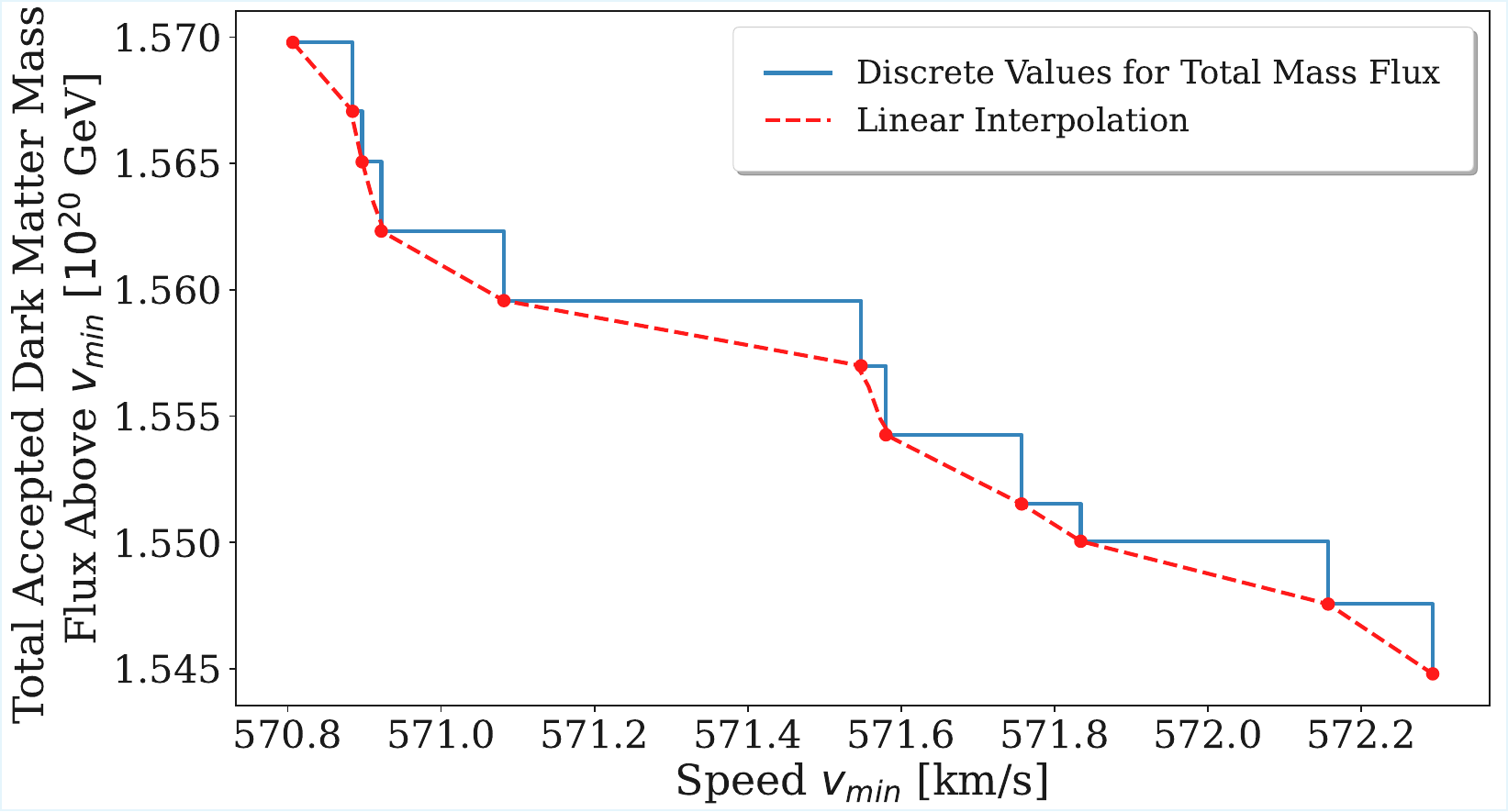}
    \caption{A closeup of the right panel of Figure \ref{fig:massFlux}, demonstrating the procedure for interpolating the flux. The solid blue line is the naive discrete counts for the flux, given by Equation \ref{eq:fluxAboveSpeed}. The red dotted line is the conservatively interpolated value which was used to compute the bounds.}
    \label{fig:ZoomIn}
\end{figure}

However, we have skipped over a subtlety of this calculation. Due to the discrete nature of the data, $N_{\geq v_0}(m_{\chi})$ is a step function of $v_0$. To compute the bounds we need to invert $N_{\geq v_0}(m_{\chi})$ to solve for $v_0$ at fixed $m_\chi$. To do so, we linearly interpolate $N_{\geq v_0}(m_{\chi})$ between the bottoms of each step in Equation \ref{eq:fluxAboveSpeed}. This procedure is shown in Figure \ref{fig:ZoomIn}.  Since this effective value for $N_{\geq v_0}(m_{\chi})$ is smaller than Equation \ref{eq:fluxAboveSpeed}, this is a conservative choice. Figure \ref{fig:massFlux} shows the interpolated value of the flux.

\section{Improved Bounds from Skylab and Ohya}
\label{sec:bounds}
In this section, we describe the improvements to the dark matter exclusion plots from plastic etch detectors at the Ohya quarry and onboard the Skylab satellite. The plastic etch detector at Ohya searched for magnetic monopoles. The other was a cosmic ray search for highly charged nuclei aboard the ISS's predecessor Skylab. Both experiments used detectors made of sheets of plastic polymer. When a particle passes through the plastic, it can damage the molecular bonds. This damage becomes visible as a hole after an acid wash. \cite{Bhoonah2021} treated a hole burrowed entirely through the detector in a straight line as the dark matter detection signal, which neither search observed. We present the relevant parameters from the experiments in Table \ref{tab:params}.

\begin{figure} 
    \centering
    \renewcommand{\arraystretch}{1.3}
    \begin{tabular}{l|c|c}  
        \centering

         & \textbf{Ohya} & \textbf{Skylab} \\
        \hline
        Area $A$ & $2442\: \text{m}^2$ & $1.17\:\text{m}^2$ \\
        \hline
        Duration $T$ & $2.1\: \text{yr}$ & $0.70\: \text{yr}$ \\
        \hline
        Tolerance Angle $\theta_{\mathrm{tol}}$ & $18.4\degree$ & $60\degree$ \\
        \hline
        Overburden Material & Atmosphere + $2.7\: \frac{ \text{g}}{\text{cm}^{3}} \text{ Crust }$  &$2.7 \:\frac{ \text{g}}{\text{cm}^{3}} \text{ Aluminum }$ \\
        \hline
        Solid Overburden Length at $\theta_{\mathrm{tol}}$ & $39 \:\text{m}$ & $0.74\: \text{cm}$ \\
        \hline
        \makecell[l]{Column Density of \\Atmospheric Overburden at $\theta_{\mathrm{tol}}$}  & $960\:\frac{\text{g}}{\text{cm}^2}$ & $0\:\frac{\text{g}}{\text{cm}^2}$ \\
        \hline
        Detector Material & $1.2\:\frac{ \text{g}}{\text{cm}^{3}} \text{ Lexan } \mathrm{(C_{16}H_{14}O_3})$ &$1.34  \:\frac{ \text{g}}{\text{cm}^{3}} \text{ CR-39 } \mathrm{(C_{12}H_{18}O_7)}$
 \\
        \hline
        Detector Length at $\theta_{\mathrm{tol}}$ & $0.66 \:\text{cm}$ &  $1.6 \:\text{cm}$ \\
        \hline
        Detection Threshold $ E'_{\mathrm{th}}$ & $0.3 \:\frac{ \text{GeV}}{\text{cm}}$ &$0.5  \:\frac{ \text{GeV}}{\text{cm}}$
    \end{tabular}

    \captionof{table}{The experimental parameters from the Skylab and Ohya experiments relevant to this work. These are taken from \cite{Bhoonah2021}.
    } \label{tab:params}
\end{figure}

The Ohya detector ran for 2.1 years. It was made of Lexan $\mathrm{(C_{16}H_{14}O_3})$ with an energy threshold of $0.3 \: \text{GeV}/\text{cm}$, a density of $1.2  \:\text{g}/\text{cm}^{3}$ and an effective area of $2442 \: \text{m}^2$. The detector had a tolerance angle of $18.4\degree$ and a thickness of $0.66\: \text{cm}$ at that tolerance angle.
Ohya's overburden consists of the entire atmosphere, then crust with a thickness of $39$ m at the tolerance angle. Information on the crustal abundances of elements was taken from the Mendeleev package in Python \cite{mendeleev2014}, which obtained its data from \cite{haynes2014crc}.
For Ohya's atmospheric overburden, we use the isothermal model, with atmospheric mass density, $\rho_A$, exponentially dependent on altitude $h$ \cite{Cappiello:2023},
\begin{equation}\label{eq:isotherm}
    \rho_A(h) = \rho_0\exp\left(-\frac{h}{h_0}\right),
\end{equation}
where $h_0 = 7\:\text{km}$ is a characteristic height and $\rho_0 = 1.3\times 10^{-3}\: \text{g}/\text{cm}^3$ is the atmospheric density at ground level. Since the tolerance angle of the detector, $\theta_{\mathrm{tol}}$, is not close to $90\degree$, the column number density of dark matter entering at the tolerance angle is $h\rho_0\sec(\theta_{\mathrm{tol}})$ to a very good approximation. We also assume that the elemental ratios of the atmosphere are independent of height with values from \cite{Bramante2023Atm}, fixed at 75.6\% nitrogen, 23.18\% oxygen, and 1.2\% argon by mass.

Now moving to Skylab, its detector made observations for 0.7 years. It was made of CR-39 $\mathrm{(C_{12}H_{18}O_7})$ with an energy threshold of $0.5 \: \text{GeV}/\text{cm}$, a density of $1.34  \:\text{g}/\text{cm}^{3}$ and an effective area of $1.17 \:\text{m}^2$. The detector had a tolerance angle of $60\degree$ and a thickness of $1.6\: \text{cm}$ at that tolerance angle.
Skylab's overburden consists of the aluminum exterior of the satellite, with a thickness of $0.74$ cm at the tolerance angle.

To compute the effective number densities in Equations \ref{eq:densPN} and \ref{eq:densCON}, we sum over all isotopes of all elements in the material. For the isotope data, we again use the Mendeleev package, which uses isotopic abundance data from \cite{Kondev_2021} and isotopic mass data from \cite{isotopeMasses}.

\begin{figure}[htbp]
    \centering
    % First row
    \begin{subfigure}[b]{0.45\textwidth}
        \includegraphics[width=\textwidth]{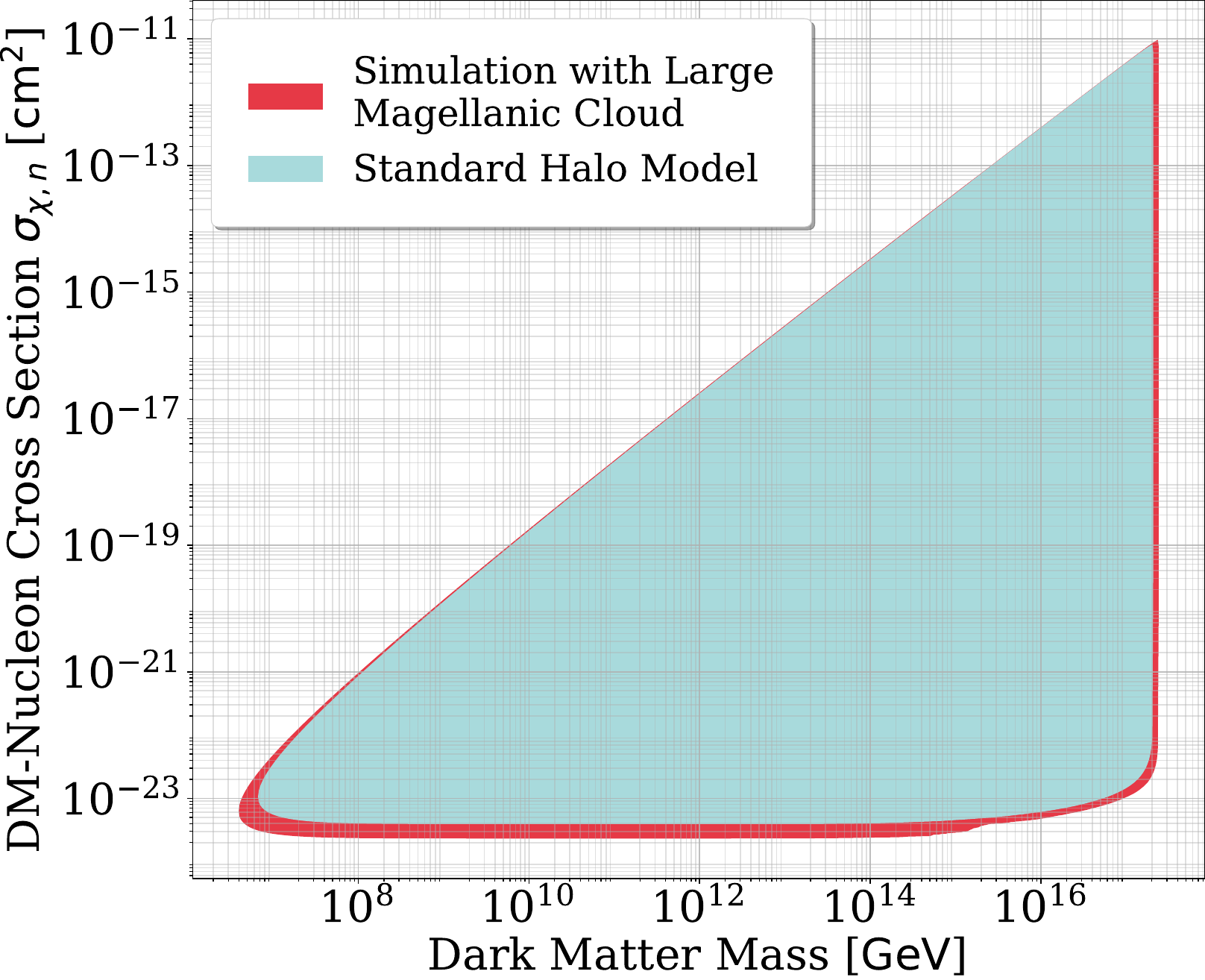}
    \end{subfigure}
    \hfill
    \begin{subfigure}[b]{0.45\textwidth}
        \includegraphics[width=\textwidth]{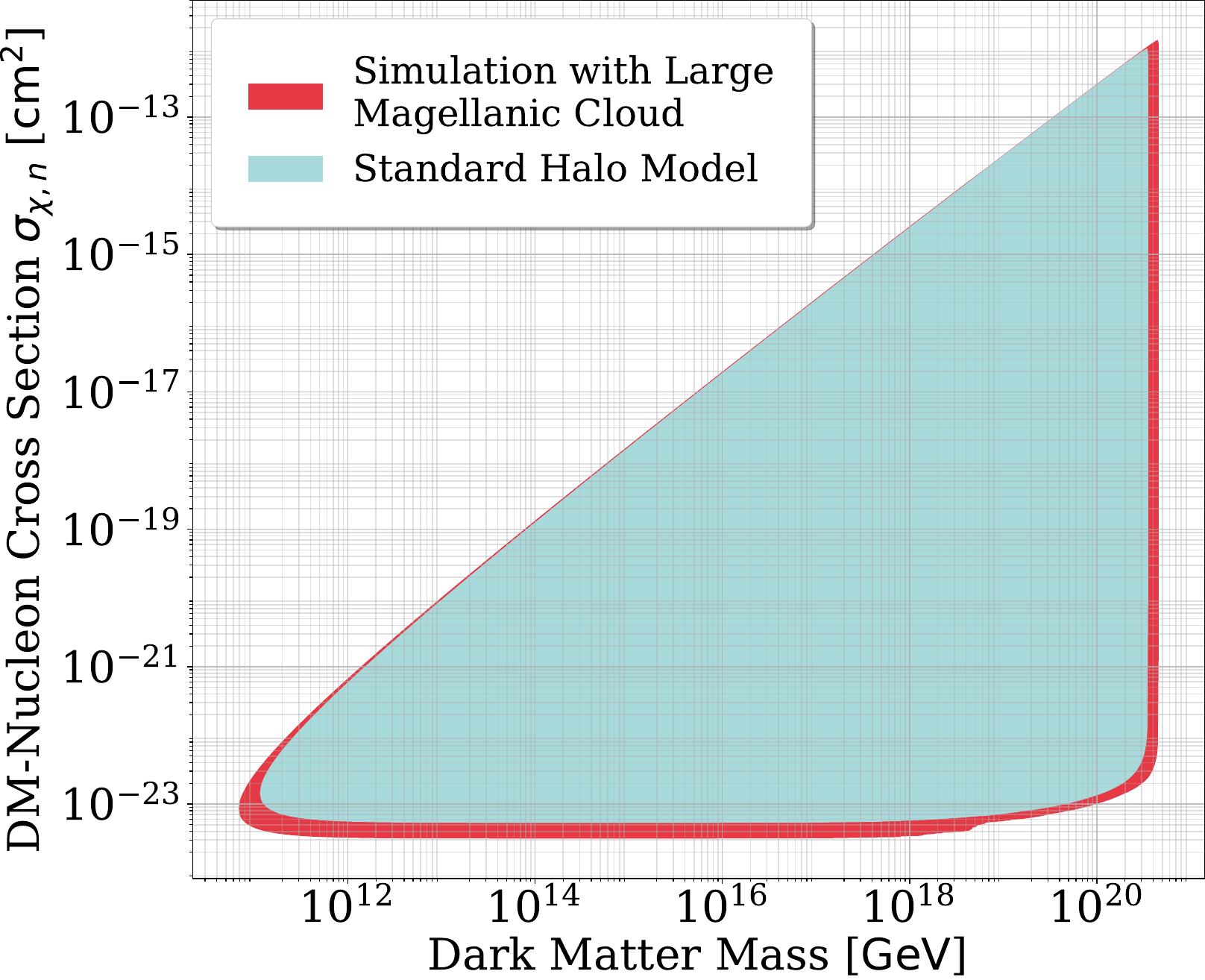}
    \end{subfigure}

    \vspace{0.5cm} % space between rows

    % Second row
    \begin{subfigure}[b]{0.45\textwidth}
        \includegraphics[width=\textwidth]{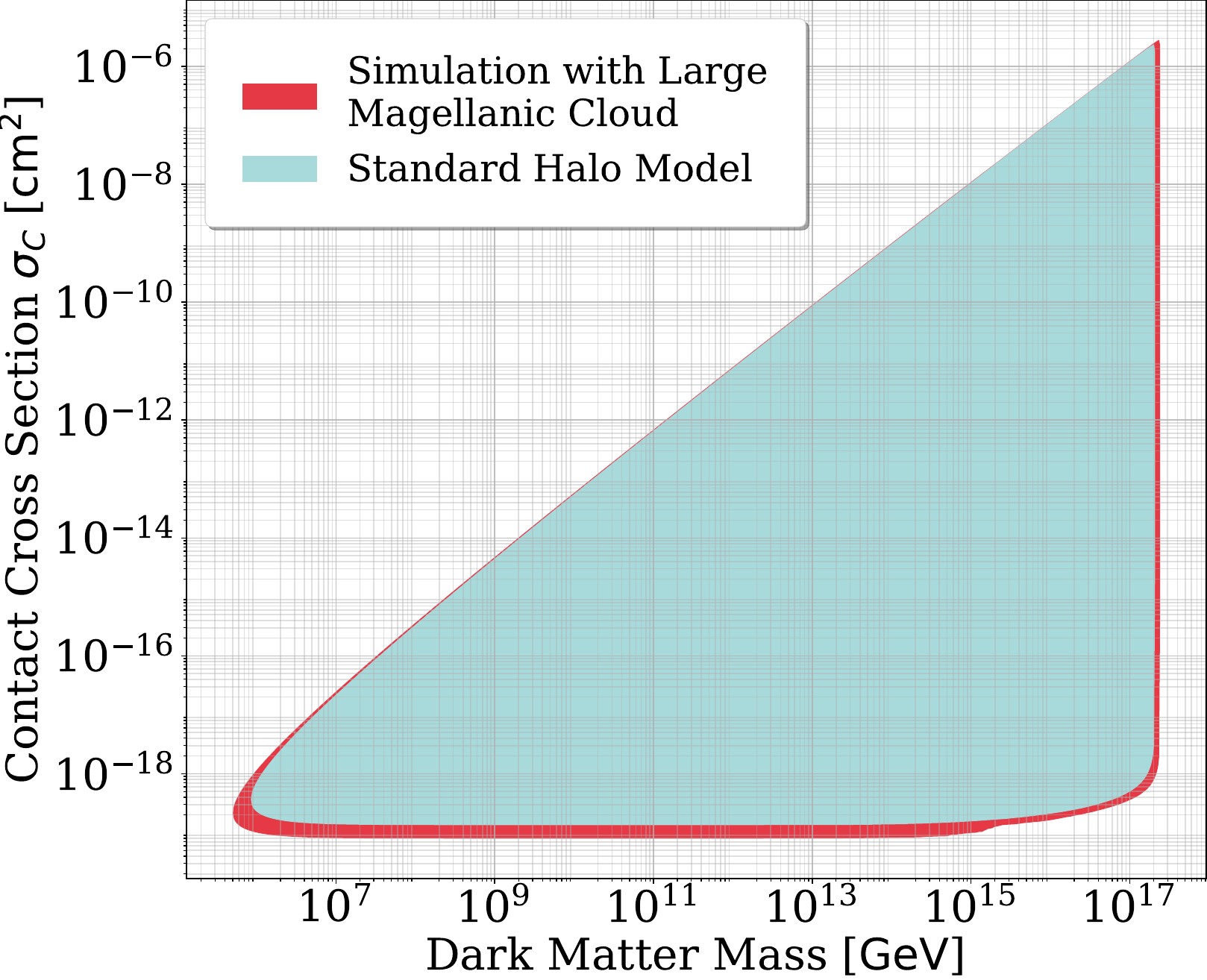}
    \end{subfigure}
    \hfill
    \begin{subfigure}[b]{0.45\textwidth}
        \includegraphics[width=\textwidth]{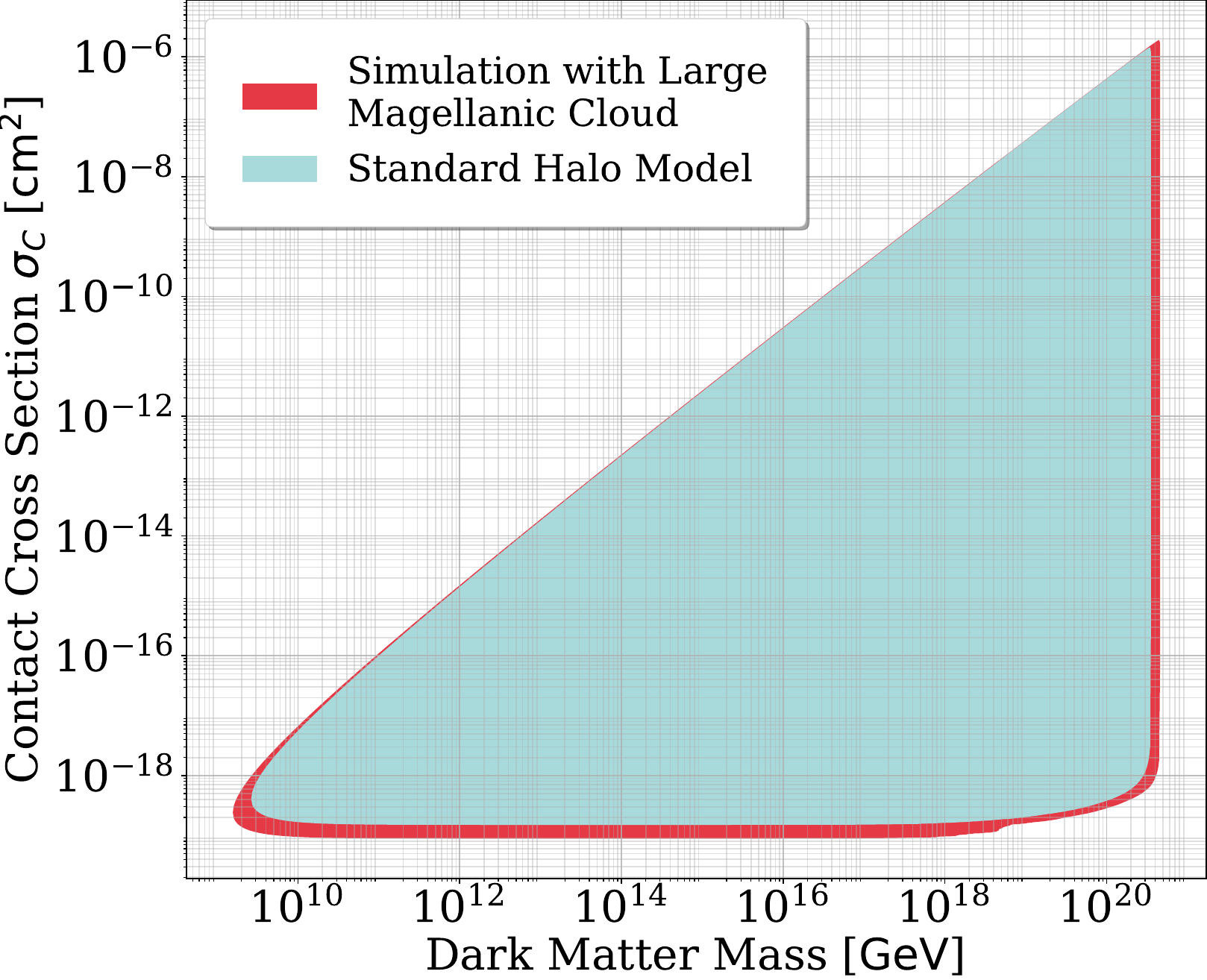}
    \end{subfigure}

    \caption{A comparison of the excluded regions from the Skylab (left) and Ohya (right) experiment assuming a per-nucleon (top) or contact (botton) interaction. The bound in light blue assumes the SHM. The bound in red uses a local dark matter velocity distribution extracted from the simulated MW-LMC analogue.}
     \label{fig:ourBounds} 
\end{figure}

Using Equation \ref{eq:bounds}, we can now compute the bounds. In Figure \ref{fig:ourBounds}, we show the bounds for the contact and per-nucleon case for both Ohya and Skylab using the data from the simulated MW-LMC analogue, shown in red. For comparison, we also recomputed the bounds using velocities drawn from the SHM, shown in blue. 

\begin{figure}[htbp]
    \centering

    % Top: Contact interaction
    \begin{subfigure}[b]{0.8\linewidth}
        \centering
        \includegraphics[width=\linewidth]{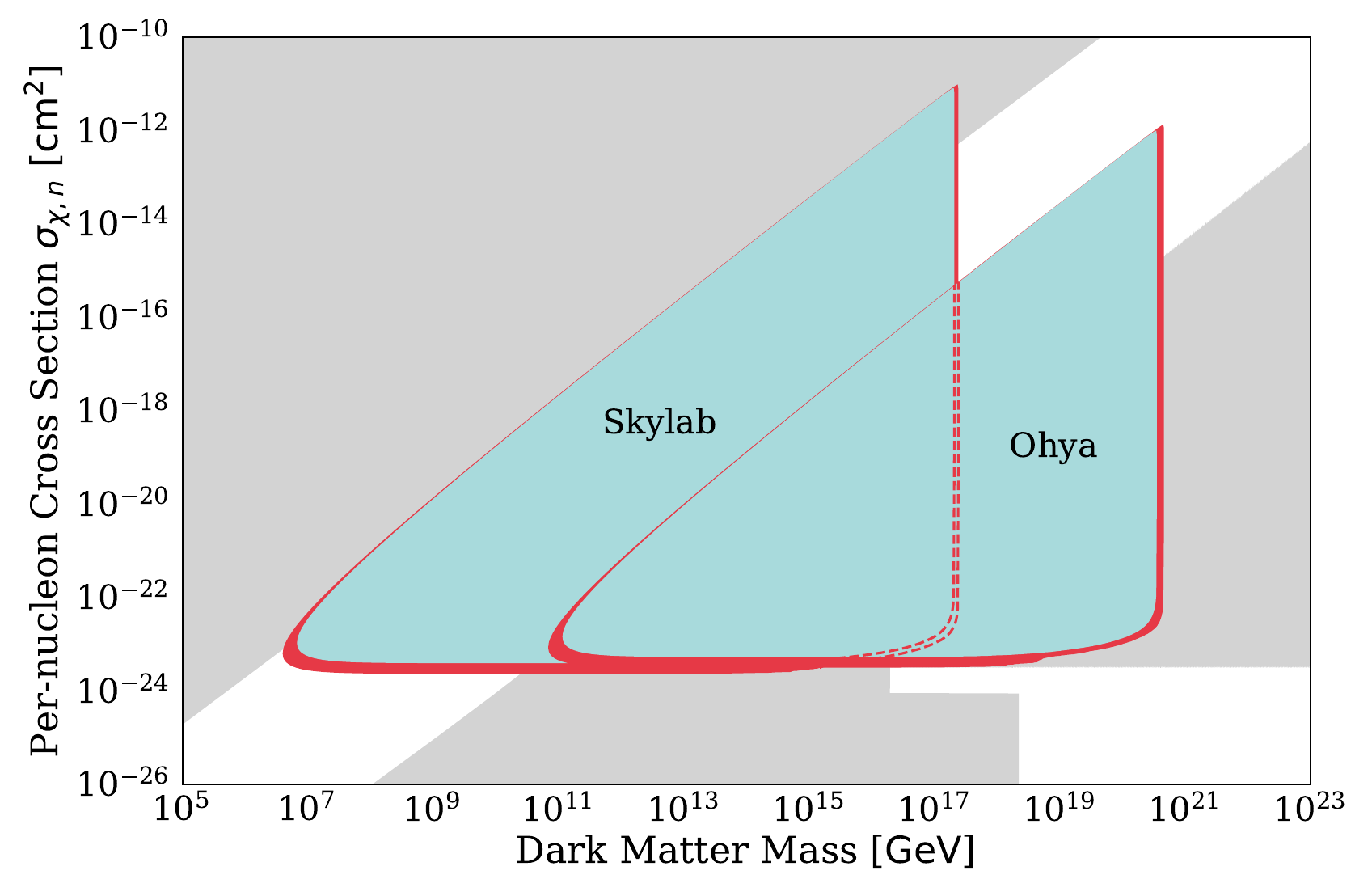}
        \label{fig:otherBounds1}
    \end{subfigure}

    \vspace{0.5cm} % space between panels

    % Bottom: Per-nucleon interaction
    \begin{subfigure}[b]{0.8\linewidth}
        \centering
        \includegraphics[width=\linewidth]{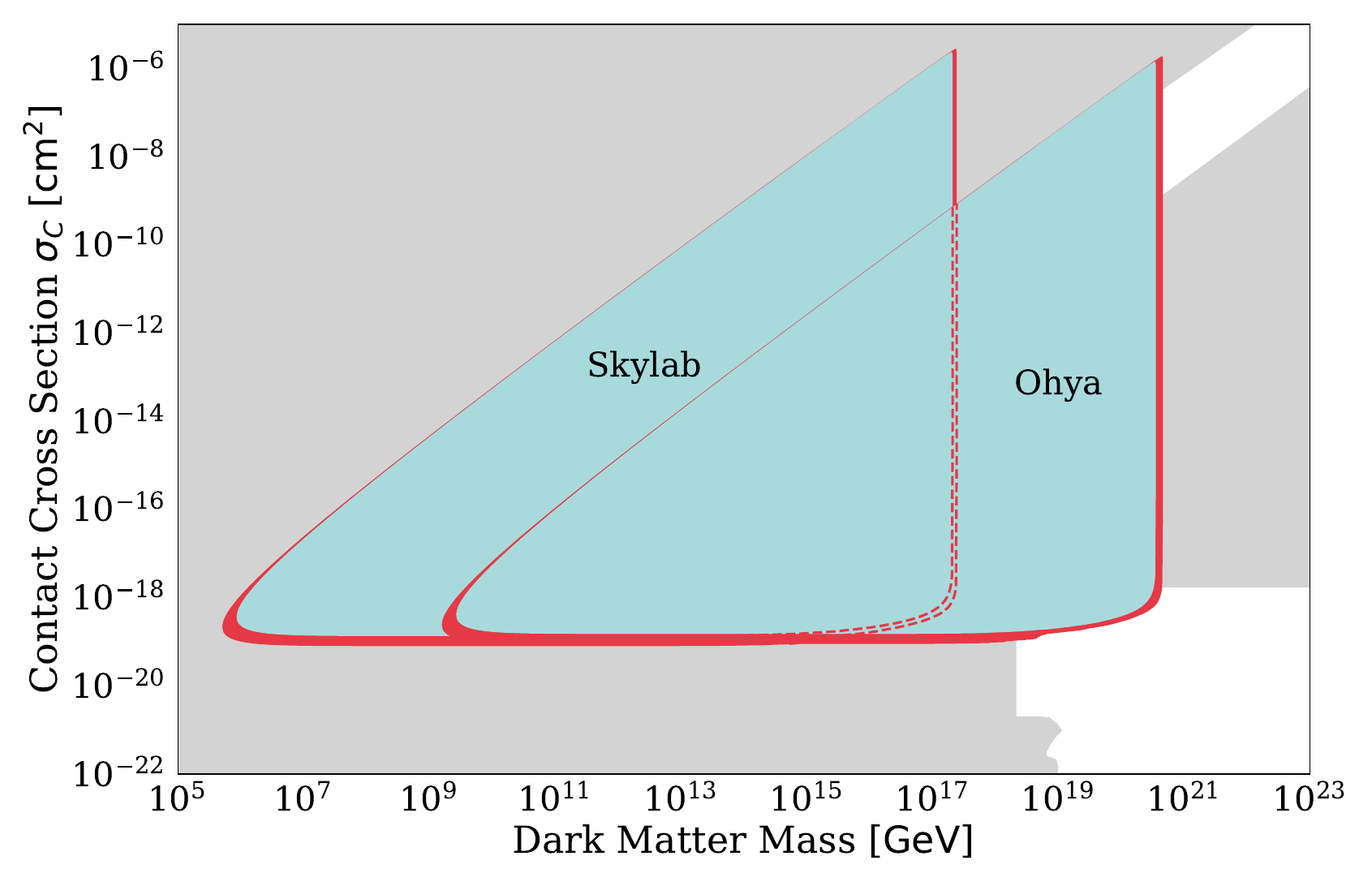}
        \label{fig:otherBounds2}
    \end{subfigure}

    \caption{The bounds from the Skylab and Ohya plastic detectors for a per-nucleon (top) and contact (bottom) interaction. In each plot, Skylab's bound is on the left and Ohya's is on the right. In light blue are the bounds assuming the SHM and in red is the additional parameter space using the results from the simulated MW-LMC analogue. These bounds are compared with other leading bounds, shown in grey.}
    \label{fig:otherBounds}
\end{figure}

We proceed with a brief discussion on the origin of the exclusion lines and the effect of the LMC on the bounds. At the high mass end, the bound stops because the flux is too small to trigger the detector. The faster particles due to the impact of the LMC increase the flux, improving the bound.
For higher cross sections, the particle loses too much energy in the overburden and can't trigger the detector. Since the loss in the overburden is exponential (see Equation \ref{eq:enegyexponential}), the faster particles due to the LMC have little effect on the bound. At small cross sections, the bound breaks down because the interaction is too weak to trigger the detector. Here, the effect of the increased flux due to the LMC is the most significant. The effect is most pronounced at low masses, where high speed particles more frequently pass through the detector. Figure \ref{fig:otherBounds} shows the result of including the effect of the LMC on Skylab and Ohya bounds when compared to leading bounds for the per-nucleon and contact cases.

\

\section{Discussion and conclusions}
In this work, we have investigated 
the effects of the LMC on heavy dark matter bounds, using a MW-LMC analogue from the Auriga magneto-hydrodynamical simulations
to
extract the local dark matter velocity distribution, and move beyond the SHM. We recalculated heavy dark matter bounds from plastic etch detectors at the Ohya quarry in Japan and aboard the Skylab satellite using the updated local dark matter halo distribution and showed that including the effects of the LMC extends the parameter space that can be probed by the experiments.

An interesting consequence of our work is that the extent of bounds derived from ground-based detectors, and their corresponding dark matter flux sensitivity, differ by latitude. Specifically, detectors placed in the northern hemisphere, at a latitude of around $30\degree$, experience the highest flux due to the Sun's motion through the galaxy. This effect is present in the SHM, but is exacerbated by the influence of the LMC, as can be seen in Figure \ref{fig:Direction Scatter}. This is a potential consideration for future ground-based heavy dark matter searches.

We note that the MW–LMC analogue used in this work is drawn from a fully cosmological simulation, which self-consistently follows the joint formation and evolution of MW–mass halos and massive satellites in a $\Lambda$CDM context. While this system is not explicitly tailored to reproduce all detailed constraints of the present day MW–LMC system, the selected LMC analogue broadly matches key observed properties, and we have further chosen the position of the Sun in the simulations to reproduce the observed Sun–LMC geometry. 

Recent studies have shown that detailed predictions for the MW–LMC interaction are sensitive to modeling uncertainties, including the poorly constrained dark matter distribution in the outer MW halo, which affects the LMC's orbit, and the initial dark matter mass and concentration of the LMC halo prior to infall, which affect the amount of tidal stripping and the resulting phase space distribution of unbound LMC dark matter particles in the Solar neighborhood (e.g.~\cite{2024MNRAS.531.3524Y, 2025MNRAS.544.1820Y, 2022MNRAS.514.1266P}). Nevertheless, by utilizing a fully cosmological simulation, our results illustrate how a massive satellite close to its pericentric passage can impact the inner halo and heavy dark matter searches. While this work focuses on a single MW–LMC analogue that reproduces key properties of the observed system, exploring a broader ensemble of MW–LMC realizations with future high resolution cosmological simulations will help quantify the associated uncertainties.

The procedure outlined in this work is applicable to any aboveground or orbital search for multiply interacting dark matter. For modern orbital searches, there will likely be sufficient information about the orientation of the detector over time, avoiding the assumption of random orientation necessary for the Skylab satellite in Section \ref{sec:flux}. We can also apply this procedure to searches for dark matter in mineral slabs, appropriately averaging over the orientation of the slab in the Earth.

\section*{Acknowledgements}

NB acknowledges the support of the Canada Research Chairs Program, as well as the Natural Sciences and Engineering Research Council of Canada (NSERC), funding reference number RGPIN-2020-07138 and the NSERC Discovery Launch Supplement, DGECR-2020-00231. JB acknowledges support from the Natural Sciences and Engineering Research Council of Canada (NSERC), the Ontario Early Researcher Award (ERA), and the Canada Foundation for Innovation (CFI). This research was undertaken thanks in part to funding from the Arthur B. McDonald Canadian Astroparticle Physics Research Institute. We thank Azadeh Fattahi for providing the re-simulated Auriga halo used in this work.

\bibliographystyle{JHEP.bst}

\bibliography{lmc.bib}

\end{document}